\documentclass[12pt]{article} 

\usepackage{latexsym} 	
\usepackage{amssymb}  	
\usepackage{amsbsy}
\usepackage{epsfig}     
\usepackage{graphics,color}
\usepackage{a4}

\newcommand{\om}{\omega}
\newcommand{\Om}{\Omega}

\newcommand{\bra}{\langle}
\newcommand{\ket}{\rangle}

\newcommand{\half}{\frac{1}{2}}

\newcommand{\vecnul}{{\mathbf 0}}  
\newcommand{\kv}{{\mathbf k}}  
\newcommand{\pv}{{\mathbf p}}

\newcommand{\xv}{{\mathbf x}}

\newcommand{\be}{\begin{equation}}
\newcommand{\ee}{\end{equation}}
\newcommand{\bea}{\begin{eqnarray}}
\newcommand{\eea}{\end{eqnarray}}
\newcommand{\bean}{\begin{eqnarray*}}
\newcommand{\eean}{\end{eqnarray*}}
\newcommand{\nn}{\nonumber}
\newcommand{\hm}{\hspace*{-0.6cm}}

\newcommand{\gm}{\gamma}

\begin{document}

\title{
\vskip -100pt
{\begin{normalsize}
\mbox{} \hfill DAMTP-2007-7\\
\mbox{} \hfill hep-ph/0701205 \\
\vskip  100pt
\end{normalsize}}
{\bf\Large Particle creation and warm inflation}
\author{
\addtocounter{footnote}{2}
Gert Aarts$^{a}$\thanks{email: g.aarts@swan.ac.uk}
 {} and
Anders Tranberg$^{b}$\thanks{email: A.Tranberg@damtp.cam.ac.uk}
 \\ {} \\
{}$^a${\em\normalsize Department of Physics, Swansea University}
\\
{\em\normalsize Singleton Park, Swansea SA2 8PP, United Kingdom}
\\ {} \\
 {}$^b${\em\normalsize DAMTP, University of Cambridge} \\
   {\em\normalsize Wilberforce Road, Cambridge CB3 0WA, United Kingdom}
}
}
\date{January 24, 2007}
\maketitle
\begin{abstract}
  During cosmological inflation, it has been suggested that fields coupled 
to the inflaton can be excited by the slow-rolling inflaton into a 
quasi-stable non-vacuum state. Within this scenario of ``warm inflation'', 
this could allow for a smooth transition to a radiation dominated Universe 
without a separate reheating stage and a modification of the slow roll 
evolution, as the heat-bath backreacts on the inflaton through friction.  
In order to study this from first principles, we investigate the dynamics 
of a scalar field coupled to the inflaton and $N$ light scalar boson 
fields, using the 2PI-$1/N$ expansion for nonequilibrium quantum fields. 
As a first step we restrict ourselves to Minkowski spacetime, interpret 
the inflaton as a time-dependent background, and use vacuum initial 
conditions. We find that the dominant effect is particle creation at late 
stages of the evolution due to the effective time-dependent mass. The 
further transfer of energy to the light degrees of freedom and subsequent 
equilibration only occurs after the end of inflation. As a consequence, 
the adiabatic constraint, which is assumed in most studies of warm 
inflation, is not satisfied when starting from an initial vacuum state.
 \end{abstract}
                                                                                
\newpage
                                                                                
                                                                                
                                                                     
\section{Introduction}
\label{sec:Introduction}
\setcounter{equation}{0}

In recent years quantitative cosmological observations have become 
available, resulting in a model for our Universe that is consistent with 
an early period of inflation and predictions of simple inflationary 
theories \cite{Spergel:2006hy}. In the most elementary setup, the dynamics 
of the inflaton mean field $\phi(t)=\bra \varphi(t,\xv) \ket$ is 
determined by
 \be 
 \label{eq:inf1}
 \ddot\phi(t) + 3H(t)\dot\phi(t) + V'[\phi(t)] = 0, 
\ee 
 subject to the Friedmann equation $H^2=(\half \dot\phi^2+V[\phi])/3M_{\rm 
Pl}^2$, where $M_{\rm Pl}$ is the Planck mass. 

In warm inflation 
\cite{Moss:1985wn,Berera:1995ie,Berera:1998px,Berera:2002sp,Moss:2006gt,Bastero-Gil:2006vr}
 the key idea is that interactions between the inflaton and other quantum 
fields are important during inflation and that they result in continuous 
energy transfer from the inflaton to these other fields. If this transfer 
is sufficiently fast and equilibration is rapid, a quasi-stable state 
could be achieved, different from the inflationary vacuum. An additional 
effective friction term would then be expected in the inflaton equation of 
motion. A simple phenomenological modification of Eq.\ (\ref{eq:inf1}) 
incorporating this reads \cite{Moss:1985wn,Berera:1995ie}
 \be 
 \label{eq:inf2}
 \ddot\phi(t) + \left[ 3H(t) + \Upsilon_\phi(t) \right] \dot\phi(t) + 
V'[\phi(t)] = 0, 
\ee 
 where $\Upsilon_\phi(t)$ is a time- and field-dependent friction 
coefficient.\footnote{For a critical analysis of warm inflation, see 
Ref.\ \cite{Yokoyama:1998ju}.}\footnote{For a study of particle 
creation and friction during power law inflation, see Ref.\ 
\cite{Yokoyama:1987an}.}

Since warm inflation necessarily involves multiple interacting quantum 
fields, which are dynamically evolving in real time, a full, quantitative 
understanding is difficult. In particular, a first-principle investigation 
requires all tools available to study quantum field dynamics far from 
equilibrium beyond the mean-field approximation. This may be contrasted 
with the theory of preheating due to parametric resonance after inflation, 
which can be understood from a combination of mean-field dynamics and the 
classical random field approximation 
\cite{Polarski:1995jg,Kofman:1997yn,Khlebnikov:1996mc}.

For the purpose of this paper, the scenario of warm inflation 
separates naturally into three parts \cite{Berera:2002sp}:

\begin{enumerate}
 \item The inflaton field $\varphi$ is coupled to a second scalar field 
$\chi$. While the inflaton rolls down the effective potential, it 
interacts with the $\chi$ field, resulting in the excitation of $\chi$ 
degrees of freedom.
 \item The field $\chi$ is coupled to other degrees of freedom, which can 
be light fermion ($\psi$) or scalar ($\sigma$) fields. These fields are 
excited as well and may thermalize. Various interaction terms are 
possible.
 \item The backreaction of the $\chi$ field on the inflaton leads to 
effective friction in the inflaton equation of motion, resulting in 
overdamped inflaton dynamics.
 \end{enumerate}

A commonly used interaction term between the inflaton field and the
second field $\chi$ is given by $\frac{1}{2} g^2\varphi^2\chi^2$. Due to
the time dependence of the inflaton $\phi(t)$, an effective
time-dependent mass term appears in the quantum dynamics of the $\chi$
field, $m_\chi^2 + g^2\phi^2(t)$, where $m_\chi$ is the mass in absence
of the inflaton. In this paper, we consider a large field model, such that
the inflaton is rolling down its effective potential and $\phi(t)$ is
decreasing. As a first step, we treat the inflaton as a time-dependent
background and assume the most extreme case, in which the slow-rolling
of the inflaton is caused by the interactions with the heat-bath rather
than the expansion of the Universe. Taking this overdamped limit with
$\Upsilon \gg H$, and assuming a quadratic potential $V[\varphi] = \half
m_\varphi^2\varphi^2$ for simplicity, the dominant solution of Eq.\
(\ref{eq:inf2}) in the overdamped regime reads $\phi(t) = \phi_0 \exp
[-(m_\varphi^2/\Upsilon) t]$. We assumed that $\Upsilon$ is time- and
field-independent.  The inflaton is treated as a dynamical field in a
companion paper \cite{inprep}.
 Here, our goal is to study the dynamics of parts 1.) and 2.) described
above, by combining a mode function analysis \cite{Birrell:1982ix} and
the techniques of the two-particle irreducible (2PI) effective action
\cite{Cornwall:1974vz} for nonequilibrium quantum fields
\cite{Berges:2004yj}. Since the inflaton acts as a time-dependent
background, we find that there is particle production in the $\chi$
sector, akin to particle production in curved spacetime
\cite{Birrell:1982ix}. In Section \ref{secmode} we study this process
using a mode function analysis. We find that details of $\chi$
particle creation depend crucially on the size of the zero-temperature
mass $m_\chi$ and the momentum $k$. Most importantly, we find that
particle production only takes place towards the end of the evolution
and that the amount of particles is, to a large extent, independent from
the initial conditions and the duration of the inflationary stage.

In order to determine the range of validity of the mode function analysis 
and study the effect of interactions, we continue in Section 
\ref{sec:inter} with a full far-from-equilibrium numerical study in 
quantum field theory using the 2PI effective action, and include 
interactions between the scalar field $\chi$ and $N$ light quantum fields. 
Specifically, we include $N$ scalar fields $\sigma_a$ ($a=1,\ldots, N$) 
and use a truncation of the 2PI effective action determined by the 
2PI-$1/N$ expansion to next-to-leading order (NLO) 
\cite{Berges:2001fi,Aarts:2002dj}. This approximation has been well 
studied in recent years and is known to give a quantitative description of 
both the early evolution far from equilibrium as well as the later stages 
of equilibration and thermalization (see e.g.\ Refs.\ 
\cite{Berges:2001fi,Aarts:2001yn,Berges:2002cz,Arrizabalaga:2004iw}).
  We consider both a trilinear coupling $\chi\sigma_a^2$ and a quartic
coupling $\chi^2\sigma_a^2$. Irrespective of the interaction term, we
find that as long as the inflaton is evolving, the mode function
analysis gives an accurate description of the dynamics in the $\chi$
sector. In particular, processes leading to equilibration and
thermalization are not yet relevant.

 In this first study, we ignore the expansion of the Universe and work 
in Minkowski spacetime. If anything, neglecting the dilution caused by
expansion should increase the chances of realising a ``warm'' state.


\section{Mode function analysis}
\label{secmode}
\setcounter{equation}{0}

As a first step, we perform a mode function analysis for a free $\chi$ 
field, subject to a time-dependent mass
 \be
 \label{eqMt}
 M_\chi^2(t) = m_\chi^2+ \delta m^2e^{-\gamma t}.
\ee  
 Here $\delta m$ contains the details on the coupling to the inflaton. 
Specifically for the $\frac{1}{2}g^2\varphi^2\chi^2$ interaction, we find 
that $\delta m^2 e^{-\gamma t} = g^2 \phi^2(t)$, but these details are not 
required here. Using the standard decomposition, we write
 \be
 \chi(t,\xv) = \int \frac{d^3k}{(2\pi)^3} \left[  
a_\kv f_\kv(t) e^{i\kv\cdot\xv} + a_\kv^\dagger f^*_\kv(t) 
e^{-i\kv\cdot\xv} \right], \ee
and find that $f_\kv(t)$ is a solution of
 \be
\ddot f_\kv(t) + \left[\kv^2 + m_\chi^2 + \delta m^2 e^{-\gamma t} \right]
f_\kv(t) =0.
 \ee
A change of variables to $x=(2\delta m/\gamma)e^{-\gamma t/2}$ shows that 
this is a Bessel equation of the form
\be
x^2f_\kv^{''}(x) + xf_\kv^{'}(x)
 + \left[ x^2 + \frac{4\om_\kv^2}{\gamma^2} \right] f_\kv(x) =0,
\ee
 where $\om_\kv=\sqrt{\kv^2+m_\chi^2}$. 
We note that only the combination $\om_\kv/\gamma$ appears in this 
equation, while the dependence on $\delta m$ will enter via the initial 
conditions.
The general solution is
\be
f_\kv(t) =
A_\kv^+ J_{2i\om_\kv/\gamma}\left(\frac{2\delta m}{\gamma}e^{-\gamma
t/2}\right)
+
A_\kv^- J_{-2i\om_\kv/\gamma}\left(\frac{2\delta m}{\gamma}e^{-\gamma
t/2}\right),
\ee
 where $J_\nu(z)$ is the Bessel function of the first kind. 
The constants $A_\kv^\pm$ are determined by the initial conditions, which 
are fixed by demanding that 
$f_\kv(0) = 1/\sqrt{2\Om_\kv}$ and $\dot f_\kv(0) = -i\Om_\kv f_\kv(0)$,
where $\Om_\kv=\sqrt{\kv^2+M_\chi^2(0)}$ and $M_\chi(0)$ 
is the initial mass. This determines $A_\kv^\pm$ as
\be
A_\kv^\pm =
\frac{\mp i}{\sqrt{2\Om_\kv}}\frac{\pi}{\gamma\sinh\left(2\pi
\om_\kv/\gamma\right)}
\left[ \dot J_{\mp 2i\om_\kv/\gamma} \left(\frac{2\delta m}{\gamma}\right) 
+ i\Om_\kv
J_{\mp 2i\om_\kv/\gamma} \left(\frac{2\delta m}{\gamma}\right)\right],
\ee
where
\be
  \dot J_{\mp 2i\om_\kv/\gamma} \left(\frac{2\delta m}{\gamma}\right) = 
\frac{d}{dt}
J_{\mp 2i\om_\kv/\gamma} \left(\frac{2\delta m}{\gamma} e^{-\gamma 
t/2}\right) \Bigg|_{t=0}.
\ee
One may verify that the Wronskian $f_\kv(t)\dot f_\kv^*(t) - \dot 
f_\kv(t) f_\kv^*(t) = i$ is preserved during the time evolution.

\begin{figure}[t]
 \centerline{\epsfig{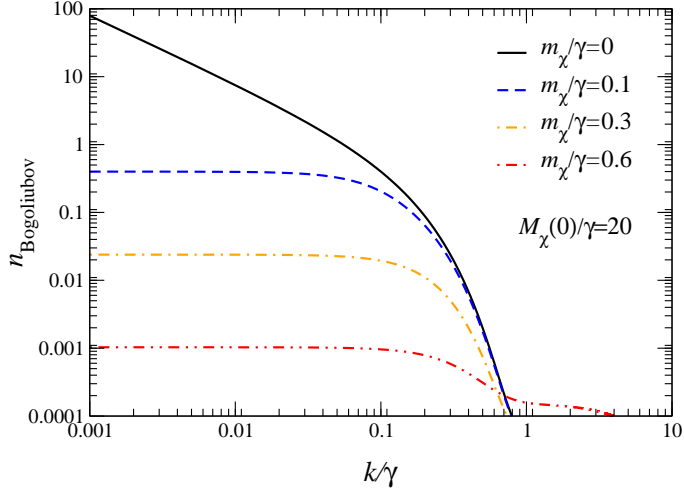}}
 \caption{Particle number $n_\kv$ as a function of $k/\gamma$ according to 
the Bogoliubov relation for four values of the final mass $m_\chi$. 
The initial mass is $M_\chi(0)/\gamma=20$.}
 \label{fig:nbog}
\end{figure}

In order to identify the produced particle number at asymptotically late 
times, we use some elementary properties of Bessel functions and find that 
\be 
 \lim_{t\to\infty}  
J_{\pm 2i\om_\kv/\gamma} \left(\frac{2\delta m}{\gamma} e^{-\gamma t/2}\right) 
= C_\kv^\pm e^{\mp i\om_\kv t},
\ee 
with
\be
C_\kv^\pm
= 
\left(\frac{\delta m}{\gamma}\right)^{\pm 2i\om_\kv/\gamma}
\frac{1}{\Gamma(1\pm 2i \om_\kv/\gamma)} 
\ee
At late times, we find therefore that the mode functions oscillate with 
the expected frequency $\om_\kv$, and   
\be 
\lim_{t\to\infty} f_\kv(t) = A_\kv^+ C_\kv^+ 
e^{-i\om_\kv t} + A_\kv^- C_\kv^- e^{i\om_\kv t}. 
\ee 
A comparison with the standard form of the mode functions at $t\to\infty$,
$\tilde f_\kv(t) = e^{-i\om_\kv t}/\sqrt{2\om_\kv}$ yields the Bogoliubov 
coefficient \cite{Birrell:1982ix}
\be
\beta_\kv = i \left[\tilde f_\kv(t)\partial_t f_\kv(t)-
f_\kv(t)\partial_t \tilde f_\kv(t)\right]
= -\sqrt{2\om_\kv}A_\kv^-C_\kv^-.
\ee
Therefore, the final particle number (starting in vacuum initially) is 
$n_\kv = |\beta_\kv|^2 = 2\om_\kv |A_\kv^-C_\kv^-|^2$,
which, after some algebra, can be written as\footnote{If the initial 
particle number is nonzero and equals $\bra a_\kv^\dagger a_\kv\ket = 
n_\kv^{(0)}$, we 
find that $n_\kv = \big( 1 + 2 n_\kv^{(0)} \big) \om_\kv |C_\kv^-|^2
 \left( |A_\kv^+|^2 + |A_\kv^-|^2 \right) - \frac{1}{2}$.} 
 \be
n_\kv =
\frac{1}{2\Om_\kv}\frac{\pi}{\gamma\sinh(2\pi \om_\kv/\gamma)}
\left\{
\left|\dot J_{2i\pi \om_\kv/\gamma}\left(\frac{2\delta 
m}{\gamma}\right)\right|^2+ \Om_\kv^2
\left| J_{2i\pi \om_\kv/\gamma}\left(\frac{2\delta 
m}{\gamma}\right)\right|^2
\right\} - \half.
\label{eq:bogonk}
\ee
The resulting particle numbers are shown in Fig.\ \ref{fig:nbog} for four 
different values of the asymptotic mass $m_\chi$ and an initial mass 
$M_\chi(0)/\gamma=20$. We observe that a significant amount of particles 
are only produced when $m_\chi/\gamma \ll 1$, and only with momentum 
$k/\gamma\lesssim 1$.

\begin{figure}[t]
 \centerline{\epsfig{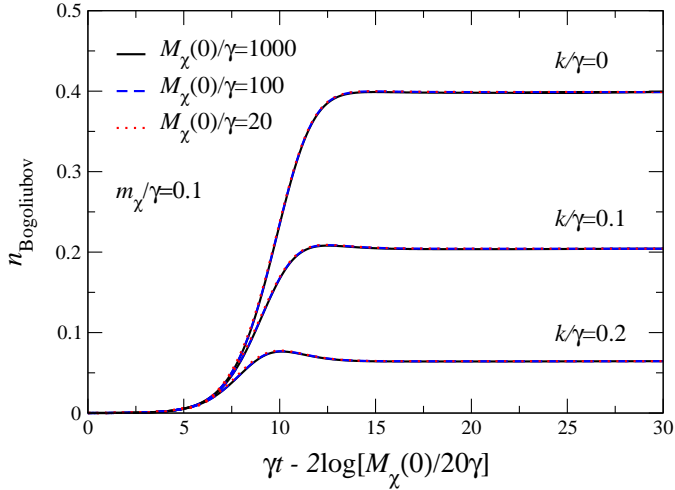}}
 \caption{Particle number $n_\kv$ as a function of time,
according to  the Bogoliubov relation, for three values of the initial mass 
$M_\chi(0)$ and three momentum modes. After shifting time 
by an amount $2\log[M_\chi(0)/M_\chi^{\rm ref}]$, where 
$M_\chi^{\rm ref}/\gamma=20$ is a reference mass, there is no 
dependence on the initial mass remaining.
The final mass is $m_\chi/\gamma=0.1$.
}
 \label{fig:nbog_t}
\end{figure}

In order to determine when particles are created during the inflationary 
stage, we show the time-dependent particle number,
 \be 
 n_\kv(t) = 
\frac{1}{2\om_\kv(t)}\left[ |\dot f_\kv(t) |^2 + \om_\kv^2(t)  | f_\kv(t) 
|^2 \right] -\half, 
 \ee 
 where $\om_\kv(t)=\sqrt{\kv^2+ M_\chi^2(t)}$, in Fig.\ \ref{fig:nbog_t}.  
We comment on possible other definitions of particle number elsewhere 
\cite{inprep}. The time-dependent particle number is shown for three 
different initial masses $M_\chi(0)$, the largest mass corresponding to 
the longest period of inflation (recall that $M_\chi(0)\sim \phi_0$).  
However, we find that a trivial shift of the time variable is sufficient 
to take the initial mass dependence into account. In other words, the 
amount of particles produced is independent of the initial conditions and 
therefore the duration of the inflationary stage. Generically, particles 
can only be produced when $|\dot\om_\kv(t)| \gtrsim \om_\kv^2(t)$, which 
in our model translates to $\gm \gtrsim M_\chi(t)$. If we denote the time 
when most particles are created with $t_*$, we find from Fig.\ 
\ref{fig:nbog_t} that $\gamma t_*-2\log[M_\chi(0)/M_\chi^{\rm ref}] 
\approx 8$, where $M_\chi^{\rm ref}/\gamma=20$ is a reference mass. The 
size of the effective mass is then $M_\chi(t_*)/\gamma\approx 0.37 < 1$, 
confirming that $\chi$ particles are only produced when $M_\chi(t)/\gamma$ 
is sufficiently small.


\section{Interactions}
\label{sec:inter}
\setcounter{equation}{0}

In the mode function analysis of the previous section, interactions 
besides the time-dependent mass are ignored. In order to assess the 
validity of that analysis, we now include interactions and consider the 
coupling of the scalar field $\chi$ to $N$ light degrees of freedom, 
either fermionic or bosonic. We start with some parametric estimates and 
then present the numerical results obtained with the help of the 2PI-$1/N$ 
expansion for nonequilibrium quantum fields.

\subsection{Parametric estimates}
\label{secpam}

In order to get an estimate for the magnitude of various parameters, we 
use here an inflaton potential $V[\varphi]=\half m_\varphi^2 \varphi^2$. 
During slow-roll inflation, the Hubble parameter is approximately given by 
$H \sim m_\varphi\phi/M_{\rm Pl}$. In order to have inflation, the initial 
amplitude should be $\phi_0\sim\, \mbox{a few} \, M_{\rm Pl}$, and to 
satisfy CMB constraints \cite{Spergel:2006hy}, we require that 
$m_\varphi\sim 10^{-6}M_{\rm Pl}$. We then find that $H \sim 10^{-6}\phi_0 
\gtrsim 10^{-6}M_{\rm Pl}$. In warm inflation, the effective damping term 
$\Upsilon$ in the inflaton equation wins over the expansion rate. Taking 
$\Upsilon\sim 100 H$ yields $\Upsilon \sim 10^{-4}M_{\rm Pl}$. Using 
exponential time dependence, $\phi(t)=\phi_0 \exp 
[-(m_{\varphi}^2/\Upsilon) t]$, then yields a value of $\gamma\sim 10^{-8} 
M_{\rm Pl}$ for the rate in the time-dependent mass (\ref{eqMt}).
 In order to have any particle production, we found from the mode
function analysis that $m_\chi/\gamma \lesssim 0.1$, or $m_\chi \lesssim
10^{-9} M_{\rm Pl}$, which means that the $\chi$ particle can be some
degree of freedom beyond the Standard Model.

We now consider the couplings to the light degrees of freedom. First 
we consider the coupling to $N$ fermion fields, with the interaction 
term
 \be 
\sum_{a=1}^{N} 
\frac{h}{\sqrt{N}}\chi\bar{\psi}_{a}\psi_{a}. 
\ee
 The factor $\sqrt{N}$ is introduced to allow for a proper $1/N$
expansion.  Following Ref.\ \cite{Berera:2002sp}, we impose $h^2
\lesssim 1$ such that perturbation theory is reasonable.\footnote{This
corresponds to $N h^2 \lesssim 1$ in the conventions of Ref.\
\cite{Berera:2002sp}.} From a standard one-loop calculation one can
compute the (zero-temperature) onshell decay width for $\chi\to
\psi\psi$, in the case that $M_\chi>2M_\psi$. Here the masses
$M_{\chi,\psi}$ include possible background field dependence. The width
is given by
 \be
\Gamma_\pv^\chi = \frac{h^2 M_\chi^2}{8\pi\sqrt{\pv^2+M_\chi^2}}
\left(1-\frac{4M_\psi^2}{M_\chi^2}\right)^{3/2}.
 \ee
 The important assumption made in most studies of warm inflation is that
the dynamics takes place in the so-called adiabatic approximation,
$|\dot\phi/\phi| \ll \Gamma_\pv^\chi$, leading to quick decay (and 
possibly
thermalization) during inflation.\footnote{The origin of the adiabatic
approximation can be traced back to Refs.\
\cite{Hosoya:1983ke,Morikawa:1985mf}.}
 In our model we have to compare $\Gamma_\pv^\chi$ with $\gamma$. Taking
for simplicity $M_\psi\ll M_\chi$ and $\pv=\vecnul$, we find
 \be
\frac{\Gamma_\vecnul^\chi}{\gamma} = \frac{h^2}{8\pi}\frac{M_\chi}{\gamma}.
 \ee
 When it is assumed that $M_\chi\sim g\phi_0$ and $\phi_0\gg \gamma$ (as
discussed above), this ratio can be much larger than one.  However, in
the mode function analysis we found that $\chi$ particles are only
produced when the (time-dependent) mass $M_\chi$ is much smaller,
specifically $M_\chi/\gamma \lesssim 1$. In that case the important
conclusion is that the dynamics is not taking place in the
adiabatic regime, but rather in the opposite limit $|\dot\phi/\phi| \gg
\Gamma_\pv^\chi$.

In the numerical study below, we couple the $\chi$ field to $N$ light 
bosonic fields, with the interaction term
\be
\sum_{i=1}^N \frac{h}{\sqrt{N}} \chi \sigma_a^2.
\ee
 In this case, the coupling constant $h$ is dimensionful and it is natural 
to write $h = m_\chi\tilde h$, where $\tilde h$ is dimensionless. We 
consider again the decay process $\chi\to \sigma\sigma$ and find for the 
onshell width, assuming that $M_\chi>2M_\sigma$,
 \be
 \Gamma_\pv^\chi = 
\frac{\tilde{h}^2 m_{\chi}^{2}}{8\pi\sqrt{\pv^2+M_\chi^2}}
\sqrt{1-\frac{4M_\sigma^2}{M_\chi^2}}.
 \ee
To test the adiabatic approximation, we compare again  
$\Gamma_\pv^\chi$ with $\gamma$. Taking $M_\sigma \ll M_\chi$ 
and $\pv=\vecnul$, we find
 \be
\frac{\Gamma_\vecnul^\chi}{\gamma} = \frac{\tilde 
h^2}{8\pi}\frac{m_\chi}{\gamma}\frac{m_\chi}{M_\chi}
\ll 1.
 \ee
 Since all factors are strictly less than one, we are not in the adiabatic 
limit. This is in agreement with the results from the numerical analysis 
carried out below.

\subsection{Nonequilibrium dynamics}
\label{sec:2pi}

To determine the range of validity of the mode function analysis and 
estimates carried out above, we continue with a numerical study using the 
2PI effective action for quantum field dynamics in real time. We refrain 
from providing details on the 2PI effective action and the 
Schwinger-Keldysh formalism for nonequilibrium field theory, instead we 
refer to Ref.\ \cite{Aarts:2002dj}, whose notation we follow closely.

We consider the following action,
\bea
S[\chi,\sigma] =
-\int d^4x &&\hm \Bigg\{ \frac{1}{2}(\partial_\mu \chi)^2 
+ \frac{1}{2} \left[ m_\chi^2 + \delta m^2 e^{-\gamma t}\right] \chi^2
+ \frac{\lambda_\chi}{4!N} \chi^4 
\nn  \\
  &&\hm
+ \frac{1}{2}(\partial_\mu \sigma_a)^2 + \frac{1}{2} m_\sigma^2 \sigma_a^2
+ \frac{\lambda_\sigma}{4!N} (\sigma_a\sigma_a)^2
+ V_{\rm int}[\chi,\sigma]
\Bigg\}.
\eea
We consider two different interaction terms $V_{\rm int}[\chi,\sigma]$ 
between the $\chi$ field and the scalar fields $\sigma_a$, a trilinear and 
a quartic coupling. Both interaction terms preserve the $O(N)$ symmetry in 
the $\sigma$ sector, such that the $\sigma$ two-point function can be 
written as ${G_\sigma}_{ab}(x,y) = \delta_{ab}G_\sigma(x,y)$.
We have scaled all couplings in such a way that a proper $1/N$ expansion 
is possible. We use $N=4$ throughout.

\underline{\em Trilinear coupling.}
 Motivated by Ref.\ \cite{Berera:2002sp}, we start with the trilinear 
coupling,
\be
 V_{\rm int}[\chi,\sigma] = \frac{h}{\sqrt N } \chi\sigma_a^2 + 
c_\chi\chi.
\ee
 Because this term breaks the symmetry $\chi\to -\chi$, we have to allow 
for a nonzero expectation value $\bar \chi(t) = \sqrt{N} 
\bra\chi(t,\xv)\ket$.\footnote{This was overlooked in Ref.\  
\cite{Berera:2002sp}.} The term linear in $\chi$ is used to shift the 
minimum of the (effective) potential at the initial time to $\bra\chi\ket=0$.
The 2PI part of the effective action is written 
as  
$\Gamma_2[G_\chi,G_\sigma] =  \Gamma_{\chi\sigma}[G_\chi,G_\sigma] + 
\Gamma_\sigma[G_\sigma]$, 
where the first term is given by the two-loop diagram
\be
 \Gamma_{\chi\sigma}[G_\chi,G_\sigma] =
  i h^2 \int d^4x\, d^4y\, G_\chi(x,y) G_\sigma^2(x,y).
\ee
The second part, $\Gamma_2[G_\sigma]$, is the standard NLO contribution 
for $N$ scalar fields. This contribution has been discussed in detail in 
Refs.\ \cite{Berges:2001fi,Aarts:2002dj} and will not be shown here 
explicitly. Other 2PI diagrams are suppressed by $1/N$ and are not 
included. 
The resulting equations of motion follow from a variation of the 2PI 
effective action with respect to $\bar \chi$, $G_\chi$ and $G_\sigma$,
and read
\be
\label{eqchi}
\frac{d^2}{dt^2}\bar\chi + 
\left[ m_\chi^2 + \delta m^2 e^{-\gamma t} +\frac{\lambda_\chi}{6} 
\bar\chi^2
+\frac{\lambda_\chi}{2N} G_\chi(x,x) \right] \bar \chi
= - h G_\sigma(x,x) - c_\chi,
\ee
and for the two-point functions ($j=\chi,\sigma$)
\be
\label{eqG}
-\left[\square_x + M_j^2(t) \right] G_j(x,y) 
= i \int d^4z\, \Sigma_j(x,z) G_j(z,y) + i \delta^4(x-y),
\ee
with the effective masses
\be
M_\chi^2(t) =  m_\chi^2 + \delta 
m^2 e^{-\gamma t} + \frac{\lambda_\chi}{2}\bar\chi^2, 
\;\;\;\;\;\;\;\;
M_\sigma^2(t) =  m_\sigma^2 + 2h\bar\chi +
\lambda_\sigma\frac{N+2}{6N} G_\sigma(x,x).
\ee
 The nonlocal contributions to the self energies $\Sigma_{\chi,\sigma}$ 
follow from variation of $\Gamma_2[G_\chi,G_\sigma]$ in the usual manner 
\cite{Aarts:2002dj,inprep}.

\underline{\em Quartic coupling.}
 In order to assess the importance of the possibility of onshell 
decay $\chi\to \sigma\sigma$ in the trilinear case, we also consider the 
following quartic potential,
 \be
 V_{\rm int}[\chi,\sigma] = \frac{h}{2N} \chi^2\sigma_a^2.
\ee
 In the 2PI effective action we include the two- and three-loop diagrams
 \be
\Gamma_{\chi\sigma}[G_\chi,G_\sigma] =
-\frac{h}{2} \int d^4x\, G_\chi(x,x) G_\sigma(x,x)
+  \frac{i h^2}{2N} \int d^4x\,d^4y\, G_\chi^2(x,y)
G_\sigma^2(x,y),
\ee
as well as $\Gamma_{\sigma}[G_\sigma]$ to NLO as above. 
Strictly speaking, we deviate here from the $1/N$ expansion, since the 
three-loop diagram only appears at next-to-next-to leading order. 
We include it nevertheless, since it plays the same 
role as the two-loop diagram in the trilinear case and results in 
$\chi\sigma$ interaction beyond the 
mean-field approximation.
 In this case the symmetry $\chi\to -\chi$ is preserved, so that we can
take $\bra \chi\ket = 0$ consistently for the entire evolution. Hence
Eq.\ (\ref{eqchi}) is absent, while the effective masses appearing in
Eq.\ (\ref{eqG}) are 
 \bea
\label{eqM1}
M_\chi^2(t) = &&\hm m_\chi^2 + \delta m^2 e^{-\gamma t} + h G_\sigma(x,x), 
\\
\label{eqM2}
M_\sigma^2(t) = &&\hm m_\sigma^2  
+ \lambda_\sigma\frac{N+2}{6N} G_\sigma(x,x)
+ \frac{h}{N} G_\chi(x,x).
\eea
 The nonlocal self energies follow again from the 2PI effective action.

\begin{figure}[t]
 \centerline{\epsfig{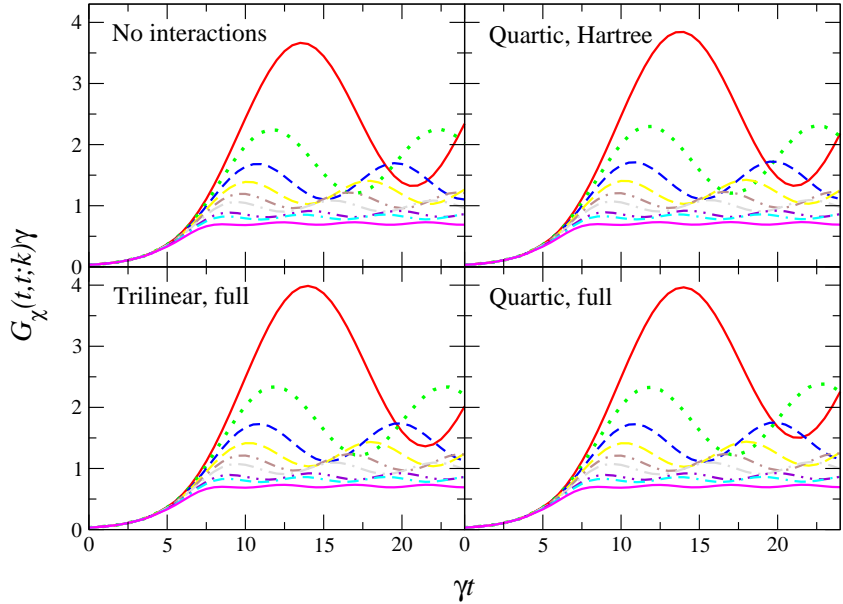}}
 \caption{Equal-time two-point function $G_\chi(t,t;\kv)$ as a function of 
time $t$ in units of $\gamma$, for nine different momenta $k$ (the 
ordering is such that the smallest momentum corresponds to the largest 
value at $\gamma t\approx 10$),
 using the analytical expression (top left), the trilinear interaction 
(bottom left), the quartic interaction term in the Hartree approximation 
(top right), and the quartic interaction term (bottom right). In all cases 
the initial mass is $M_\chi(0)/\gamma=20$ and the final mass is 
$m_\chi/\gamma=0.1$.
 }
 \label{fig:Gchi}
\end{figure}

We solve the resulting equations numerically, following the approach in
\cite{Berges:2001fi,Aarts:2001yn,Berges:2002cz,Arrizabalaga:2004iw}.
Space is discretized on a three-dimensional lattice with $32^3$ sites
and a physical size of $\gamma L=32$. The length of the memory kernel is
$\gamma t=20$, containing 800 time steps. The masses are
$m_\chi/\gamma=0.1$ and $m_\sigma/m_\chi=1/3$. The coupling constants
are $\lambda_\chi=6$, $\lambda_\sigma=6$. In the trilinear case the
coupling $h/m_\chi=5/3$, while in the quartic case we use $h=1$. We
initialize the two-point functions in vacuum. Renormalization is carried
out in such a way that the set of equations is initialized at the fixed
point of the renormalized mean field equations
\cite{Aarts:2000wi,inprep}.

\begin{figure}[t]
 \centerline{\epsfig{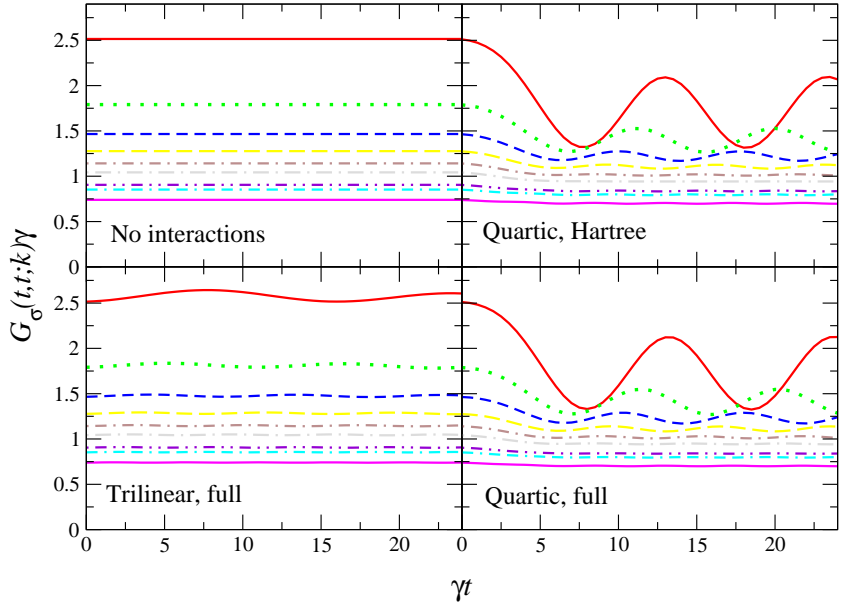}}
 \caption{Same as in Fig.\ \ref{fig:Gchi}, for the $\sigma$ two-point 
function $G_\sigma(t,t;\kv)$. The largest value at $t=0$ corresponds 
to the smallest momentum, 
since $G_\sigma(0,0;\kv)=\half(\kv^2+m_\sigma^2)^{-1}$.
 }
 \label{fig:Gsigma}
\end{figure}

Particle number is a derived concept and not always well defined in an
interacting field theory out of equilibrium. Instead of comparing
time-dependent particle numbers, we prefer to study basic quantities
appearing in the dynamical equations: the equal-time two-point functions
$G_\chi(t,t;\kv)$ and $G_\sigma(t,t;\kv)$. In Figs.\ \ref{fig:Gchi} and
\ref{fig:Gsigma} we show the evolution of the nine lowest momentum modes
in time (the evolution of the zero momentum mode is not shown). In the
top left corners, the dynamics without interactions is presented, in
which case $G_\chi(t,t;\kv) = |f_\kv(t)|^2$ and
$G_\sigma(t,t;\kv)=\half(\kv^2+m_\sigma^2)^{-1}$. The latter is exactly
time-independent, because of our choice of initialization. In the other
three frames, we show the dynamics with trilinear interaction (bottom
left), and with quartic interaction using the Hartree approximation,
i.e.\ including the self-consistently determined masses (\ref{eqM1})
and (\ref{eqM2}) but not the nonlocal terms (top right), and the full
evolution (bottom right).

As can be seen in Fig.\ \ref{fig:Gchi}, the evolution of $G_\chi$ is in
all cases nearly identical to the free case, except for a small
additional growth in the case that the nonlocal self energies are included.
From this we conclude that the mode function analysis describes the
important aspects of this part of the dynamics extremely well, even in
the presence of interactions. In particular, the process of particle
production can be studied in a satisfactory manner using the techniques
of Section \ref{secmode}.

The response of the $\sigma$ propagator is shown in Fig.\ 
\ref{fig:Gsigma}.  Recall that without interactions, the equal-time 
propagator is constant in time. In the trilinear case interactions with 
the $\chi$ field have a small effect, in the low-momentum modes only and 
with a characteristic time scale much longer than $1/\gamma$. In 
particular, there is no notion of equilibration and thermalization. This 
is consistent with the estimates given above and also with previous 
studies of nonequilibrium scalar field dynamics using the 2PI effective 
action \cite{Berges:2001fi,Berges:2002cz,Arrizabalaga:2004iw}, in which it 
was shown that thermalization occurs only after a time of the order of at 
least several hundred elementary oscillations.
 In the quartic case, the situation is slightly different because of the
presence of the $\chi$ tadpole in the effective $\sigma$ mass
(\ref{eqM2}).  During the evolution, the mass $M_\sigma$ grows as
$G_\chi(x,x)$ increases due the time-dependent inflaton background. The
response of the $\sigma$ propagator is captured very well in the Hartree
approximation. Additional nonlocal diagrams (Fig.\ \ref{fig:Gsigma},
bottom right) play only a minor role at this stage, since in the theory
with a quartic coupling onshell decay $\chi\to\chi\sigma\sigma$ is
kinematically not allowed, while scattering processes
$\chi\chi\to\sigma\sigma$ are suppressed due to small amount of $\chi$
particles present.


\section{Outlook}
\label{secout}
\setcounter{equation}{0}

In order to investigate the dynamics of warm inflation, we studied the 
quantum evolution of a scalar field $\chi$ coupled to the inflaton as well 
as $N$ light scalar fields. In this first analysis we treated the inflaton 
as a background field and ignored the expansion of the Universe. We 
took vacuum initial conditions. 
From a 
comparison with nonequilibrium quantum field dynamics using the 2PI-$1/N$ 
expansion to next-to-leading order, we found that the response in the 
$\chi$ sector can be accurately understood from a mode function analysis 
and that further interactions with light scalar fields are subdominant. 
The important conclusion is that $\chi$ particle production only takes 
place towards the end of the evolution, independent of the duration of the 
inflationary stage. An immediate consequence is that under these 
conditions the so-called 
adiabatic approximation, $|\dot\phi/\phi| \ll \Gamma^\chi$, assumed in 
most studies of warm inflation, is violated. Therefore, the dynamics 
during inflation takes place far from equilibrium and processes important 
for thermalization become important only later. These findings are 
complementary to those obtained in Ref.\ \cite{Yokoyama:1998ju}.

Although our investigation used a specific $\varphi^2\chi^2$ interaction 
and assumed an exponential time dependence for the inflaton background 
field, we believe that most of the results obtained here are generic 
for systems initially in vacuum. In particular, $\chi$ particles can only 
be produced when the rate of time variation of the effective mass is 
comparable with the effective mass itself. This observation is a potential 
hurdle in all models where the effective $\chi$ mass receives a 
contribution from interactions with the inflaton and the inflaton 
initially has a large expectation value (large field models). Concerning 
the time dependence, we have also studied a linear evolution, $\phi(t)\sim 
t$, and found that the specific time dependence is not crucial 
\cite{inprep}.

As a next step, we plan to treat the inflaton as a dynamical quantum
field, and use 2PI effective action techniques to analyse the effective
inflaton equation of motion, including the backreaction from $\chi$
particles.  Since we found in this paper that the role of the additional
light degrees of freedom is subdominant, we can focus entirely on the
inflaton-$\chi$ sector.

                                                                                
\vspace*{0.5cm}
\noindent
{\bf Acknowledgments.}
 We thank Arjun Berera for motivating us to carry out this work and Jan 
Smit, Arjun Berera, Rudnei Ramos and Ian Moss for comments.
 G.A.\ is supported by a PPARC Advanced Fellowship. A.T. is supported by 
the PPARC SPG ``Classical Lattice Field Theory''. This work was partly 
conducted on the COSMOS Altix 3700 supercomputer, funded by HEFCE and 
PPARC in cooperation with SGI/Intel, and on the Swansea Lattice Cluster, 
funded by PPARC and the Royal Society.


\end{document}